\documentclass[12pt,aps,pre,onecolumn,showpacs,superscriptaddress,showkeys,groupedaddress,nofootinbib]{revtex4-1}
\usepackage{graphics}
\usepackage{amsmath}
\usepackage{graphicx}
\usepackage{color}
\usepackage{comment}
\usepackage{lineno}
\begin{document}

\title{Synchronization transitions in spiking networks with adaptive coupling}
\author{A. Provata}
\affiliation{Institute of Nanoscience and Nanotechnology, National Center for Scientific Research ``Demokritos'', 15341 Athens, Greece}
\author{G. C. Boulougouris}
\affiliation{Institute of Nanoscience and Nanotechnology, National Center for Scientific Research ``Demokritos'', 15341 Athens, Greece}
\affiliation{Department of Molecular Biology and Genetics, Democritus University of Thrace, 68100 Alexandroupolis, Greece} 
\author{J. Hizanidis}
\affiliation{Institute of Nanoscience and Nanotechnology, National Center for Scientific Research ``Demokritos'', 15341 Athens, Greece}
\affiliation{Institute of Electronic Structure and Laser, Foundation for Research and Technology-Hellas, 70013 Herakleio, Crete, Greece}

%
%

%
\date{Received: \today / Revised version: date}
%

\begin{abstract}
\centerline{\bf Abstract}
Adaptive link sizes is a major breakthrough step in evolving networks and
is now considered as an essential process both in biological and artificial neural networks.
In adaptive networks the link weights change in time and, in brain dynamics, these
changes are controlled by the potential variations of the pre- and post-synaptic neurons.
In particular, in biological neural networks the adaptivity of the links (synapses) was 
first addressed by D. Hebb who proposed the rule that neurons which fire together wire together.
In the present study, we explore the effects of adaptive linking in networks where
hybrid synchronization patterns (solitaries, chimeras and bump states) are observed in the absence of 
adaptivity (i.e., for constant coupling strengths). The network consists of Leaky Integrate-and-Fire (LIF)
neurons coupled nonlocally in a 1D ring geometry with periodic boundary conditions. The adaptivity
follows the Hebbian principle adjusted by the Oja rule to avoid unbounded increase of the coupling
strengths. Our results indicate that, for negative coupling strengths,
the adaptive LIF network may transit through chimera state regimes with different multiplicity, 
provided that the time scales governing the link dynamics are relatively small compared
to the time scales of the potential evolution. Moreover, the size of the coherent and incoherent domains
of the chimera and their multiplicity change following the average coupling strength in the network.
Similar results are shown for the case of bump states;
for positive coupling strengths, the adaptive LIF network transits through states of different number of
active and subthreshold domains. Here also, the size and number of active and subthreshold domains
follow the evolution of the average coupling strength.
Such transient effects are suppressed when the time scales governing the link adaptive dynamics become of
the same order as the time scales governing the potential dynamics.
From the point of view of applications, adaptive networks are relevant 
 in brain dynamics and functions such as learning processes,
long and short term potentiation, gradual synapses growth in infancy and adolescence, or synapse deterioration due to
aging, neurodegenerative disorders and traumas.

\end{abstract}


\maketitle

%

\section{Introduction}
\label{intro}

\par Adaptive networks have recently attracted the attention of the scientific community
because of their applications in Biology, Sociology, Economics and Technology
 as well as in AI, a domain which is currently in
full expansion \cite{rabinovich:2006,berner:2023}. Moreover, the developments in AI find their roots in the field
of evolving neural networks, since the first AI models were composed by idealized neurons
connected in large networks organized in many layers, while the learning processes rely
on the adaptivity of the connections between the different neurons and layers. Therefore, whether
we consider biological or artificial neural networks, their interconnections must not be
assumed static but dynamic (variable), depending on various factors
including the potential environment of the 
pre- and post-synaptic elements \cite{hebb:1949}. Based on these considerations, we investigate
in this study the 
influence of adaptive connectivity in networks which present hybrid synchronization
patterns in the absence of adaptivity.
More precisely, we impose adaptive coupling in networks of coupled nonlinear neuronal oscillators
which exhibit bump states or chimera states when the link
weights are constant in time and identical in magnitude.
Such states (solitaries, chimeras and bumps) have been intensively investigated 
in the past two
decades because they support hybrid, inhomogeneous steady states 
even if the networks are composed of
identical and identically coupled elements. The aim of the present study is to highlight
the effects of adaptivity on such networks, to show that the networks may cross over
between different chimera and bump state regimes, and to investigate
the role that the time scale of the adaptation process plays in the network evolution.

\par Previous studies on chimeras and bumps without adaptation, have demonstrated the
presence of these hybrid synchronization states in a variety of nonlinear oscillator
networks, such as in networks composed by Kuramoto \cite{kuramoto:2002,abrams:2004,oomelchenko:2018},
 FitzHugh-Nagumo \cite{omelchenko:2013, omelchenko:2015},  Hindmarsh-Rose
 \cite{hizanidis:2014}, van der Pol  \cite{ulonska:2016,nganso:2023}
or
 Quadratic Integrate-and-Fire oscillators\cite{laing:2001,oomelchenko:2024} and many others.
In particular, in the Leaky Integrate-and-Fire (LIF) model 
bump states and chimera states have been reported in networks with excitatory and inhibitory coupling strengths,
respectively \cite{tsigkri:2016,tsigkri:2017,avitabile:2023,provata:2025b}. These results have been confirmed 
in the LIF model in one, two and three dimensional geometries/dimensions 
\cite{tsigkri:2017,schmidt:2017,provata:2025,kasimatis:2018}. All the aforementioned results are
produced under constant uniform coupling strengths.

\par Hybrid synchronization patterns find possible applications in the first-night effect (FNE),
an effect recorded in humans during traveling. In the first night away from home,
one hemisphere of the brain remains more alert than the other, functioning as a nocturnal sentinel 
tracking unfamiliar activity during sleep \cite{tamaki:2016}. Other unihemispheric sleep 
effects have been reported for migratory birds and sea mammals (seals and dolphins). 
During long distance flights, migratory birds sleep with one eye open 
to maintain awareness of their flight path \cite{rattenborg:2006}.
Similarly,  seals and dolphins keep one eye open during rest to monitor the surroundings
 for potential danger \cite{rattenborg:2000}.
In all these examples, the hemisphere  associated with the open eye operates in a distinct 
state of synchronization compared to the hemisphere linked to the closed eye, 
even though the two hemispheres remain interconnected as usual \cite{ramlow:2019}.

\par Regarding adaptive dynamical schemes in coupled oscillator networks, recent studies
investigate whether hybrid synchronization states can be still observed if we consider 
a co-evolving process between the coupling
strengths (and/or the topology of connections) with the node potentials.
In most cases, the studies have been conducted
using the Kuramoto model and to a lesser extent the FitzHugh-Nagumo model. 
The evolution of the couplings mostly depends on trigonometric functions of
the phase variable $\theta_j$, $j=1, \cdots N$, having frequently the form
$\frac{\mathrm{d}\sigma_{jk}}{\mathrm{d}t} \sim \cos (\theta_k-\theta_j-a )$, where $\sigma_{jk}$
is the coupling weight between nodes $j$ and $k$, $a$ is a
constant and $N$ is the number of oscillators in the network. 
The trigonometric form in the equation governing the coupling evolution is
important for keeping the coupling ranges finite. Such studies have 
concluded that the
network dynamics together with the adaptive couplings 
  can induce complex synchronization properties including chimera states
\cite{gutierrez:2011,aoki:2011,kasatkin:2017,chandrasekar:2014,huo:2019}. 

\par In the recent literature, adaptive coupling has been proposed as a key mechanism
for synaptic modification in brain dynamics and is thought to underlie learning 
and memory processes. Adaptivity was first suggested by D. O. Hebb in his 1949 book 
``The Organization of Behavior'' and it introduces local learning rules
in the sense that each synapse is updated locally. This means that the
connectivity between two specific neurons depends only on the 
activity (potentials) of the
pre- and post synaptic neurons and not on the overall brain activity.
Additionally, the gradual updating of the local connectivity 
due to the local environment and/or external stimuli
is considered as a learning process \cite{hebb:1949}.
It must be noted that direct application of the Hebbian rule in neuronal dynamics leads to
diverging coupling strengths, which is not biologically acceptable.
To address the issue of unbounded coupling growth, several modifications
 of the Hebbian learning rule have been proposed.
The most frequently used ones are the covariance rule
and the Oja rule. The covariance learning rule is
an idea similar to mean-centering of the data: by subtracting their mean values when updating
the coupling weights, the resulting differences can be either positive or negative.
It is then possible to stabilize the coupling growth using this covariance rule.
On the other hand, the Oja rule not only constrains the coupling weights to remain 
finite but also drives them toward a stable steady state, thereby 
enabling the system to adapt and learn over time \cite{oja:1982,oja:1989}.
The Oja rule will be used in this study for the updating of the network
weights to allow them to develop in a slow or fast pace
depending on the chosen timescale parameters.

\par In the next sections, we present numerical studies of LIF neuron
networks under adaptive nonlocal coupling. Starting from random initial
potentials we show evidence that a) for slowly adapting couplings, the network
crosses chimera states and bump states of different multiplicity while
for b) fast adapting couplings, the network rapidly converges into
the final state without demonstrating intermediate hybrid states.
Our findings indicate that the partial synchronization states 
emerging in neural networks by setting
appropriate connectivity schemes and/or initial conditions
are not
artificially constructed  phenomena,   but they
appear, survive, and develop  even under conditions of coupling variations and
adaptivity. 

\par
The work is organized as follows: 
In the next section, Sec. \ref{sec:model}, we present the adaptive version of the leaky Integrate-and-Fire
model, where the links evolve according to the Oja rule of Hebbian plasticity.
More specifically, in Sec.~\ref{sec:model-uncoupled} we briefly recapitulate the uncoupled LIF neuron, while in Sec.~\ref{sec:model-coupled} we introduce the coupled dynamics on the network.
The network is a 1D ring (periodic boundary conditions), whose nodes are nonlocally connected. 
The links of the network are allowed to vary following Hebbian plasticity rules.
 In Sec.~\ref{sec:quantitative},
we propose quantitative measures, such as the average link weight and the Kuramoto order parameter, 
 that are used as different measures of synchrony in the system.
In Sec.~\ref{sec:effects-adaptive}, we demonstrate various transitions between different organization
regimes as the average link weight increases from negative to positive values. In the same section,
we also study the opposite case where the link weights decrease from positive to negative values.
 In Sec.~\ref{sec:time-scales}, we study the effects of time scales and show that the different 
synchronization regimes appear only in the slow time scales, where the system has the time
to relax and visit the different synchronization states during the temporal integration. 
In Sec.~\ref{sec:discussion}, we present complementary results on the local evolution of the coupling
strengths and their distributions in space and time.
In the concluding section, Sec.~\ref{sec:conclusions}, we summarize our main results and discuss open problems.

\section{The LIF network with nonlocal interactions and adaptive coupling}
\label{sec:model}
Integrate-and-Fire (IF) models are of the oldest models proposed to capture the dynamics of neural cells.
Various types of IF models have been proposed in the literature, depending on the particular applications.
The different IF variants include the ones with linear or 
periodic or homogeneous/inhomogeneous input, 
stochastic IF models, the perfect integrator model, the quadratic, the exponential and the
leaky IF models, to name just a few 
\cite{lapicque:1907,lapicque:1907b,lapicque:1907a,burkitt:2006,santos:2019,luccioli:2010,olmi:2010,olmi:2019b,politi:2015,politi:2018,ullner:2020}.
IF models are very popular amongst computational neuroscientists due to their simplicity while retaining
the main features of neuronal activity. In the present study, we  use the 
leaky Integrate-and-Fire (LIF) model which includes a leakage term
prohibiting the potential to grow to arbitrarily large values \cite{gerstner:2002}.
In the following, we first describe the dynamics of a single LIF neuron, in Sec.~\ref{sec:model-uncoupled},
while the adaptive nonlocal coupling is introduced in Sec.~\ref{sec:model-coupled}.
Quantitative indices to monitor the evolution of the couplings and the overall synchronization
 in the system are proposed in Sec.~\ref{sec:quantitative}.

\subsection{The uncoupled LIF model}
\label{sec:model-uncoupled}

\par The mathematical description of the uncoupled LIF model is fairly simple: 
it consists of one differential equation accounting for the potential integration
and one event-driven algebraic condition accounting for the potential resetting.
It reads \cite{tsigkri:2017}:
\begin{subequations}
\begin{equation} 
 \frac{du(t)}{dt}= \mu-u(t)  
\label{eq01a} 
\end{equation}
\begin{equation}
 \lim_{\epsilon \to 0}u(t+\epsilon ) \to u_{\rm rest},  \>\>\> {\rm when} \>\> u(t) \ge u_{\rm th} 
\>\> \boldsymbol{\mid} \> {\rm with} \>\> u_{\rm th} < \mu .
\label{eq01b}
\end{equation}
\label{eq01}
\end{subequations}
\noindent 
The first equation, Eq.~\eqref{eq01a}, describes the integration of the membrane potential, $u(t)$, of the neuron.
Without the leakage term, $-u(t)$, the potential would grow linearly to infinity, with growth rate
equal to $\mu$. The introduction of the leakage term, $-u(t)$, drives the system to an asymptotic fix point
$u(t\to\infty ) =\mu$ (in the absence of algebraic condition Eq.~\eqref{eq01b}). 
Condition~\eqref{eq01b} accounts for the resetting of the potential to its ground state value, $u_{\rm rest}$,
before the potential reaches its fixed point. Namely, condition~\eqref{eq01b} states
that the potential resets to its rest state value every time  $u(t)$ reaches 
the threshold potential $u_{\rm th}$. The threshold potential needs to respect the inequality 
$u_{\rm th} < \mu$. In the opposite case ($u_{\rm th} \ge \mu$), the potential keeps increasing slowly 
and it asymptotically tends to its limiting value $\mu$ before ever reaching the value $u_{\rm th}$.

\par Starting from an initial potential $u_0$, 
the single (uncoupled) neuron described by Eqs.~\eqref{eq01a} and~\eqref{eq01b} behaves as a relaxation oscillator.
Between the initial state $u_0$ and the threshold value $u_{\rm th}$, the solution of the system, Eqs.~\eqref{eq01},
is analytically given as: $u(t)=\mu -(\mu-u_0)e^{-t}$. The period $T_\mathrm{s}$
 of the single neuron can be obtained
if the system, Eqs.~\eqref{eq01}, is integrated from $u_{\rm rest}$ to $u_{\rm th}$ and is calculated as:
$T_\mathrm{s}=\ln \left[ (\mu -u_{\rm rest})/(\mu - u_{\rm th})\right] $. $T_\mathrm{s}$ takes always finite positive
values, because both $u_{\rm rest}$ and $u_{\rm th}$ are always less than $\mu$.

\par Using the formula for the period of the single LIF neuron, we can calculate the corresponding firing rate.
Because the LIF oscillator resets once every time the potential reaches the value $u_{\rm th}$ and because
the firing  is set to take place at the resettings, the firing rate, $f_\mathrm{s}$,
of the single LIF is identical to the inverse period, as:
\begin{eqnarray}
 f_\mathrm{s}=\frac{1}{T_\mathrm{s}}=\frac{1}{\ln \left[ (\mu -u_{\rm rest})/(\mu - u_{\rm th})\right]}.
\label{eq02}
\end{eqnarray}

\par In natural conditions, the neural cell undergoes a period of refractoriness $T_\mathrm{r}$, after reseting
\cite{gerstner:2002,ermentrout:1998}.
This means that after resetting the neuron remains unresponsive to external inputs
or inputs from other neurons for a period $T_\mathrm{r}$. The value of $T_\mathrm{r}$ is an integral feature
of the type of neuron and, in most cases, it  accounts roughly for half of the period of the neural oscillations. 
To keep the system as simple as possible, in the present study, we do not consider 
refractoriness and, therefore, we set $T_\mathrm{r}=0$.

\par Equations~\eqref{eq01} and the  discussion thereafter
 describe the dynamics of a single, uncoupled LIF
neuron. In the following sections, coupling is introduced between
neurons in the 1D ring-network with adaptive, nonlocal linking.

\subsection{The coupled LIF network with adaptive nonlocal coupling}
\label{sec:model-coupled}

\par We now introduce the coupled LIF network which consists of
a (1D) ring containing $N$ nodes with LIF elements situated on 
all nodes. The state of the system at time $t$ is 
given primarily by a vector potential $\Bigl( u_1(t), u_2(t), \cdots , u_j(t), \cdots , u_N(t)\Bigr)$, $j=1, \cdots N$. 
 The LIF elements are nonlocally linked and each element is connected
symmetrically with $R$ elements to its left and $R$ elements to its right.
This type of coupling can be roughly related to the gap-junction 
(or electrical synapse) mechanism\footnote{Chemical synapses represent
a different exchange mechanism between neuron cells and are not taken
into account in the present study.}.
Periodic boundary conditions
are assumed throughout the system integration to avoid finite size effects
and to ensure that all
elements have the same number of connected elements ($2R$) and, therefore,
all elements in the system perform under the same conditions.
The coupling strengths $\sigma_{jk}(t)$ between any nodes $j$ and $k$ is time-dependent, adaptive to the
pre- and post-synaptic potentials, and their time evolution
is governed by Hebbian learning rules as modified by E. Oja \cite{hebb:1949,oja:1982,oja:1989}.
\par More specifically, assuming common parameters
$( \mu, u_{\rm rest}, u_\mathrm{th} )$ to all network nodes,
the network dynamics is described by the following two equations and one algebraic condition:
\begin{subequations}
\begin{equation}
 \frac{du_{k}(t)}{dt}= \mu - u_{k}(t) +\frac{c_u}{2R}\sum_{j=k-R}^{k+R}\sigma_{jk}(t)
[u_{j}(t)-u_{k}(t)] 
\label{eq03a} 
\end{equation}
\begin{equation}
 \lim_{\epsilon \to 0}u_{k}(t+\epsilon ) \to u_{\rm rest},  \>\>\> {\rm when} \>\> u_{k}(t) \ge u_{\rm th} 
\>\> \boldsymbol{\mid} \> {\rm with} \>\> u_\mathrm{th} < \mu
\label{eq03b}
\end{equation}
\begin{equation}
\tau_{\sigma}\frac{d\sigma_{jk}(t)}{dt}=u_j u_k -\alpha u_j u_j \sigma_{jk}.
\label{eq03c}
\end{equation}
\label{eq03}
\end{subequations}

\par Equation~\eqref{eq03a} represents the evolution of the potential of element $k$,
which follows LIF dynamics and is interacting with all nodes $j$ in the region $k-R \le j \le k+R$.
$R$ is called the coupling range and is identical for all nodes.
The interactions follow a nonlocal diffusive mechanism and interaction (coupling) strengths
between nodes $k$ and $j$ are time dependent, $\sigma_{jk}(t)$. The last term in Eq.~\eqref{eq03a},
which represents the nonlocal interactions of node $k$, is normalized over all coupled units $(2R)$.
A control parameter $c_u$ is inserted to regulate the overall strength of the coupling of potentials
 and is identical for all neurons. Therefore, the overall effective coupling strength, $\sigma^{\mathrm{eff}}_{jk}$,
corresponding to the
potential exchanges between nodes $j$ and $k$ is:
\begin{eqnarray}
\sigma^{\mathrm{eff}}_{jk}(t)=c_u \> \sigma_{jk}(t) .
\label{eq04}
\end{eqnarray}
\par The algebraic resetting condition, Eq.~\eqref{eq03b}, is identical to the corresponding one
in the case of the uncoupled neuron, Eq.~\eqref{eq01b}. Namely, the neurons reset when their 
potentials reach the threshold value $u_{\rm th}$.
We may note here that, although the condition Eq.~\eqref{eq01b} is identical for all nodes, each node 
in the network may fire
at a different time instance, e.g., in cases of incoherent networks or in the case of neighbor oscillators
belonging both to the same incoherent domain.

\par Equations~\eqref{eq03a} and \eqref{eq03b} have been studied in the past and have produced
interesting hybrid synchronization phenomena for constant values of $\sigma_{jk}$. Namely,
chimera states \cite{olmi:2010,tsigkri:2016,olmi:2019,schmidt:2017,kasimatis:2018} and
bump states \cite{tsigkri:2017,provata:2024}
have been reported in 1D rings as well as in 2D and 3D networks of LIF elements.
In the present study,
Eq.~\eqref{eq03c} is added to allow for the evolution of the coupling strengths $\sigma_{jk}(t)$.
The first term on the right hand side of this equation, $u_j u_k$, assimilates the Hebbian
rule which states that ``neurons which fire together wire together''.
If both $u_j$ and $u_k$ take high values, the corresponding coupling strength $\sigma_{jk}$
has the tendency to increase proportionally. Even in the case of small coupling strengths
the $\sigma_{jk}$ values increase slightly. As a result, if only the first term is retained
in Eq.~\eqref{eq03c},
the coupling strengths increase without bounds since the potentials in the LIF model
take always positive (or zero) values. To cure this abnormal increase, E. Oja introduced a
balancing term proportional to $-\alpha u_j u_j$ which acts as a compensation to the
perpetual increase of the coupling strength and is known as the ``forgetting term'' \cite{oja:1982,oja:1989}. 
With the addition of the last term, the average coupling strength becomes finite
as the system reaches its asymptotic equilibrium
state, 
$\left< \sigma_{jk} \right> =1/\alpha$,
while $\left< u_j u_k \right> = \left< u_j^2 \right>$, for all $j$ and $k$. 
The constant $\tau_\sigma$ represents the time scales where the evolution
of $\sigma_{jk}$ takes place compared with the time scales of the potential evolution.
We will see in the later sections that if the values of $\tau_\sigma$ are large, the
changes in $\sigma_{jk}$ are slow and the system has the chance to slowly transit through,
realize and explore many different synchronization states, while in the opposite case the system
rapidly enters the final asymptotic
state. The control parameters $\tau_\sigma$ and $\alpha$ are also identical for all elements. 
As stated earlier, periodic boundary conditions apply to all  nodes
$(j,k)$, i.e., all indices  in Eq.~\eqref{eq03} are considered $\mod N$.

\subsection{Quantitative measures}
\label{sec:quantitative}

In this section, we present collectively the  measures that
are used for the quantifications of transitions as the system
crosses between the different states of synchrony. Such measures
concern mainly the properties of the coupling matrix,
the asymptotic distribution of coupling strengths,
 the population average/mean of the coupling strengths
and its evolution in time, $\sigma (t) $,
the instant firing measures of all elements, $ f_{j}(t)$, 
the Kuramoto order parameter, $r(t)$,
and their spatial or temporal distributions.

\par Consider the coupling strength between nodes $j$ and $k$, $\sigma_{jk} (t)$.
Due to the coupling adaptivity $\sigma_{jk} (t)$ is time-dependent and is also
different for each $(j,k)$ pair. The average coupling strength, $\sigma(t)$,
and the average effective coupling strength over 
the network, $\sigma^{\mathrm{eff}}(t) $,  are defined as: 
\begin{subequations}
\begin{equation}
 \sigma(t) = \frac{1}{N(N-1)} \sum_{j,k,\> j\ne k}^{N}\sigma_{jk} (t)
\label{eq05a} 
\end{equation}
\begin{equation}
 \sigma^{\mathrm{eff}}(t) = c_u\> \sigma (t).
\label{eq05b} 
\end{equation}
\label{eq05}
\end{subequations}
Note that $\sigma(t)$ and $\sigma^{\mathrm{eff}}(t)$ are time-dependent because the population average in Eq.~\eqref{eq05a} is
over the network nodes and not over time. Throughout the rest of this study, the effective values of the couplings, 
$\sigma^{\mathrm{eff}}_{jk} (t)$ and $\sigma^{\mathrm{eff}} (t)$, 
will be used since the product of the control value $c_u$ with the 
coupling weights is the quantity which drives the coupling terms in Eq.~\eqref{eq03a}.

\par The firing rate concerning the single LIF neuron has been discussed
in Sec.~\ref{sec:model-uncoupled} and Eq.~\eqref{eq02}. However, in the
case of neurons interacting in a network the different neurons fire at
different instances, especially in the case of asynchronous neurons.
Furthermore, in the case of adaptive networks, where the coupling strengths
change in time following the potential variations, the rates also become
time dependent and the ``average firing rate'' of a neuron becomes obsolete. 
Only the instantaneous firing frequency of a specific neuron $j$, $f_j(t)$,
can be defined as follows. During the simulation, the instances when neuron
$j$ spikes are recorded as $T_j^1,\> T_j^2, \> T_j^3, \> \cdots , T_j^{m-1},\> T_j^{m},\cdots$,
where $T_j^m$ denotes the time $t$ where the neuron $j$ has emitted its $m-$th spike.
Then the instantaneous firing measure $f_j(t)$ can be approximated as : 
\begin{equation}
 f_j(t) =  \frac{1}{T_j^{m}-T_j^{m-1}}, \>\>\>{\rm for}\>\>\> T^{m-1}_j < t\le T^{m}_j .
\label{eq06} 
\end{equation}
The instantaneous firing measure of a single neuron may not provide important information
of the overall behavior of the system, but firing distributions may be more 
appropriate to describe the state of the network at specific time instances. As usual,
the firing distribution $P(f,t)$ is defined as the (normalized) ratio of neurons 
displaying firing measures in the interval $[ f, f+df )$.

\par A well known measure of synchrony in any network is the Kuramoto order parameter, $r$,
which is defined using the instantaneous phases of all network oscillators \cite{strogatz:2000,kuramoto:2002}. 
In the LIF network, the instantaneous phase $\theta_j (t) $ of oscillator $j$ 
 is defined in terms of the instantaneous potential $u_j(t)$ and varies between 0 and $2\pi$. 
The order parameter $r$ is defined as:
\begin{align}
&\theta_j = \frac{2\pi u_j}{u_{\rm th}} \nonumber \\ 
& Z = \frac{1}{N} \sum_{j=1}^{N}{\rm e}^{i \, \theta_j}   \\ 
& r=|Z|.  \nonumber 
\label{eq09} 
\end{align}
\noindent $r$ takes it maximum value, $r=1$,
when the system is in synchrony and all oscillators operate in phase. $r$ takes its minimum value, $r=0$,
when the system is in full asynchrony and all phases are randomly distributed between 0 and $2\pi$. 
At intermediate  values, $0 < r < 1$, the network demonstrates mixed behavior, acquiring both synchronous
and asynchronous oscillators. For this reason, the parameter $r$ is also often called the phase coherence
of the network and can be used as a synchrony measure in any network composed of interacting oscillators.

\par As working parameter set for this study we use the following values:
$\mu=1.0$, $u_{\rm th}=0.98$, $u_{\rm rest}=u_0=0$,
$R=350$ (see also $R=200$ in the Supplementary Material, Sec. I), $N=1024$, $\alpha=1.0$, while we mainly vary the parameters
governing the evolutions of the links, $\tau_{\sigma}$ and $c_u$.
All simulations start from random initial potentials, $u_k(t=0)$, randomly, uniformly
and independently distributed between $u_{\rm rest}$ and $u_{\rm th}$.

\section{Effects of adaptive coupling}
\label{sec:effects-adaptive}
In this section we present some first examples of transitions that the LIF network
undergoes as the average coupling changes, first  
from positive (excitatory) effective values
to negative (inhibitory) ones, in  Sec.~\ref{sec:positive-negative}, while in Sec.~\ref{sec:negative-positive} we consider 
the opposite case.

\subsection{Adaptive transitions from positive to negative effective coupling values}
\label{sec:positive-negative}
\begin{figure}[h]
\begin{center}
\includegraphics[height=0.84\textwidth]{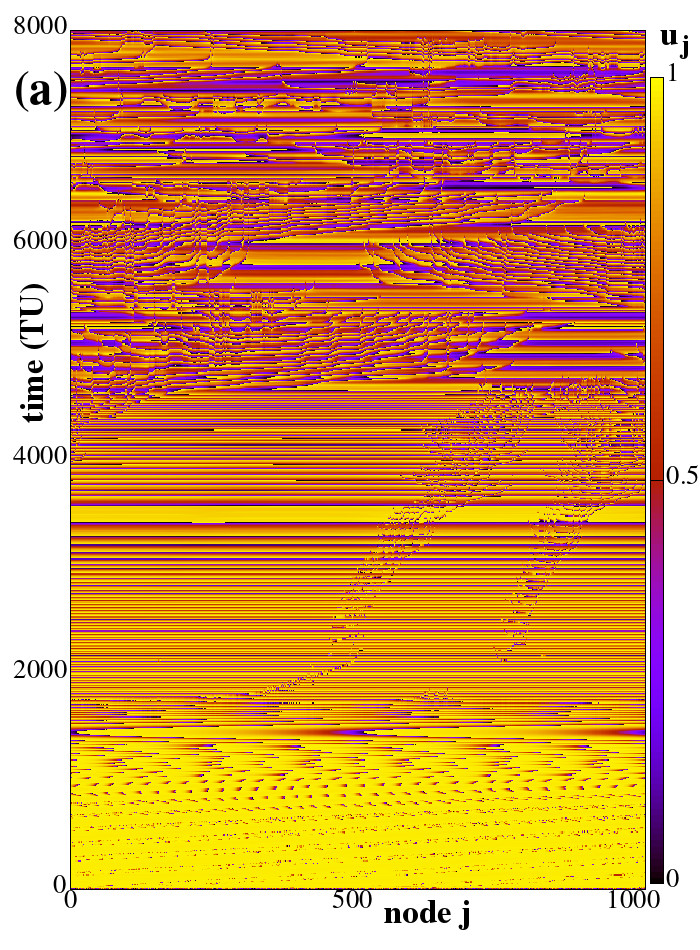}\hspace{3.1mm}
\includegraphics[height=0.89\textwidth]{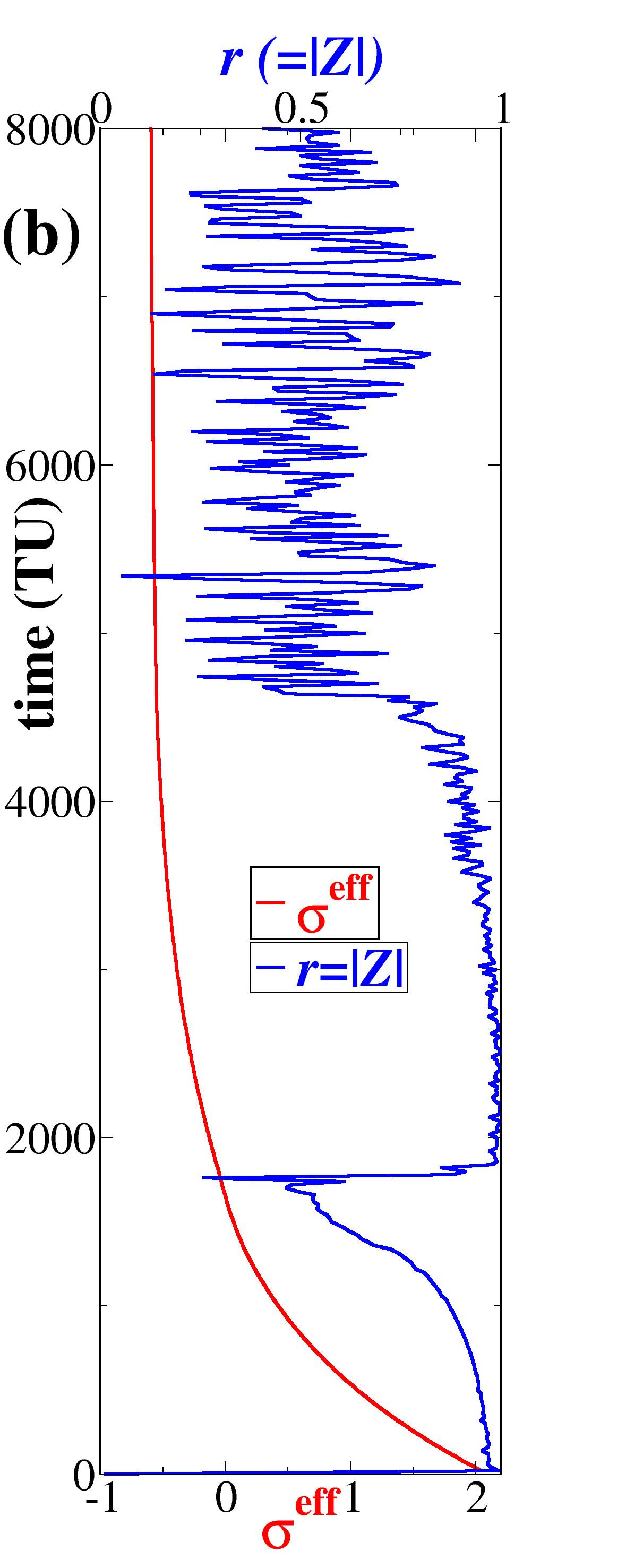}
\end{center}
\caption{\label{fig01}
(Color online) (a) Spacetime plot of the LIF network under adaptive dynamics and (b) the average effective 
coupling strength $\sigma^{\mathrm{eff}}$ (in red color) and the Kuramoto order parameter $r$
(in blue color) as a function of time. Note that $\sigma^{\mathrm{eff}}$
 decreases, starting from +2.1 at $t=0$ and approaches asymptotically
the $c_u (=-0.7)$ value. Other parameters are: $\mu=1.0$, $u_{\rm th}=0.98$, 
$u_{\rm rest}=u_0=0$,
$R=350$, $N=1024$, $\alpha=1.0$, $\tau_{\sigma}=1000$ and $c_u=-0.7$.
All simulations start from random, uniformly distributed initial potentials.
}
\end{figure}

To demonstrate multiple synchronization transitions as a result of adaptive coupling dynamics,
we plot in Fig.~\ref{fig01}a the spacetime plot for the case where the system of Eqs.~\eqref{eq03}
starts from random and homogeneously distributed initial potentials,
such that  $u_j(t=0)\in [0,u_{\rm th}]$ for all $j=1 \cdots N$. 
We have initially selected all coupling strengths $\sigma_{jk}(t=0)$ constant in value and equal in magnitude, 
while they quickly get differentiated due to the coupling dynamics governed by Eq.~\eqref{eq03c}.

\par In this section, we first want to start from positive effective coupling strengths,
where we know that bump states are observed, and via the 
adaptation process to cross over to the negative  effective couplings,
where we know that chimera states are found. The evolution of the coupling weights, $\sigma_{jk}$,
is governed by Eq.~\eqref{eq03c} and, as was discussed in Sec.~\ref{sec:quantitative}, starting from
any initial state $\{ \sigma_{jk}(t=0)\}$, the weights will finally converge asymptotically 
to $\sigma_{jk}(t\to\infty)=1/{\alpha}$. Equivalently, starting from
effective couplings $\{ \sigma^{\mathrm{eff}}_{jk}(t=0)\} =\{ c_u\> \sigma_{jk}(t=0) \}$, the 
effective weights will finally converge asymptotically 
to $\sigma^{\mathrm{eff}}_{jk}(t\to\infty)=c_u/{\alpha}$. 

\par As an example, at $t=0$ we select a relatively high and negative value of $\sigma_{jk}(t=0)=-3.0$, 
which multiplied by the negative control parameter
$c_u (=-0.7)$ produces positive effective initial couplings, $\sigma^{\mathrm{eff}}_{jk}(t=0)=c_u \> \sigma_{jk}(t=0)=2.1$.
Such positive effective couplings place initially the network in the regime of bump states.
Using as Oja constant $\alpha =1$ in Eq.~\eqref{eq03c}, we expect that in the asymptotic state the 
coupling weights will converge to values of the order $\sigma_{jk}(t\to \infty)=c_u/\alpha=-0.7$.
The negative effective couplings that the system is expected to reach asymptotically, drive
 the network toward the regime of chimeras as time increases. To allow the system enough time
to explore intermediate synchronization regimes,
we use a high value of the time scale parameter determining the coupling evolution, $\tau_{\sigma}=1000$,
in order to force the coupling process to occur in much slower scales than the potential integration.

\par The results in Fig.~\ref{fig01}a demonstrate that at $t=0$ the system starts from random potential
values, while it rapidly changes to bump states composed by coexisting subthreshold (yellow) domains intercepted
by active (brown-red) domains traveling to the right. The traveling direction (right or left) is determined
by the specific initial conditions. While it is hard to discern the transition from the random initial potentials
to the traveling bump state, this is easier done by observing the Kuramoto order parameter in Fig.~\ref{fig01}b,
blue line. At $t=0$, the Kuramoto order parameter takes the value $r(t=0)=0$ indicating full incoherence,
while subsequently $r$ jumps to a value close to $r=1$, which shows that most of the potentials take very close
values (are in coherence). 
Indeed, in bump states many nodes stay subthreshold taking similar values very close to $u_{\rm th}$, while
a few active bumps cross the system. The red line in Fig.~\ref{fig01}b indicates that at $t=0$ the 
system starts with an average effective coupling strength close to $\sigma^{\mathrm{eff}}_{jk}(t=0)=2.1$, 
as explained in the previous paragraph.

\par As time advances, the active bumps grow in sizes and, therefore, the value of the Kuramoto order parameter
decreases, since the active bumps are composed by incoherent elements. We also note a change in the direction
and composition of the active bumps with time.
At the same time, the values of the effective coupling weights decrease accordingly.  
When the effective coupling reaches the value 0, around $t=1800$ TU,  the Kuramoto order parameter shows an
abrupt drop toward $r\to 0$, because all elements perform independently in the absence of couplings.
The order parameter does not hit precisely the value $r=0$ because, just before the time when $\sigma^{\mathrm{eff}} =0$
occurs, there is a degree of coherence in the system due to the presence of the subthreshold elements.

\par Directly after the transition about $\sigma^{\mathrm{eff}}=0$, the system enters negative effective coupling 
strengths and for these parameter
values it is known that chimera states can be supported. Indeed, in Fig.~\ref{fig01}a 
and for times $1800 < t < 4300$
the system develops first a one-headed and soon after a two-headed chimera state. 
At the beginning the incoherent regions have small sizes
and as time advances their sizes grow. The appearance of the thin chimera state is
captured by the variations of the Kuramoto order parameter which immediately increases toward $r\to 1$, because
most of the system operates coherently and only the two slim incoherent regions demonstrate a certain randomness
in their structure. The broadening of the coherent regions, as time increases, is reflected by the corresponding
values of $r$ which drop slightly. Around $t=4300-4400$ TU, the two incoherent domains merge forming a single-headed
chimera which travels to the right. At this point, the coupling weights are closely approaching their asymptotic
values, $\sigma^{\mathrm{eff}}=-0.7 \>\> (\alpha=1)$. The single chimera survives for about 2000 TU and it then reduces to
disorganized coherent and incoherent domains moving around the ring-network. This disorganized phase is
captured by the large fluctuations in the Kuramoto order parameter which fluctuates away from full synchrony
($r=1$) but is also away from full asynchrony $(r=0)$. 

\par  The above discussion shows the rich synchronization
transitions caused by the Hebb-Oja adaptation process in 
neural networks, as demonstrated by the spacetime plot in Fig.~\ref{fig01}a
and the corresponding plot of the Kuramoto order parameter.
It is interesting that, although the Kuramoto order parameter captures the abruptness
of the system transitions, the evolution of the coupling weights is smooth.
Further details on the evolution of the coupling matrix $\sigma_{jk}(t)$ will be provided in Sec.~\ref{sec:discussion}.

\par In the Supplementary Material, Sec. I, we also provide spacetime plots demonstrating 
similar synchronization transitions
for other parameter values, e.g. $R=200$.
In the next section, we study the inverse scenario, where the system starts in the parameter regions
where chimera states
are produced ($\sigma^{\mathrm{eff}} < 0$) and evolves toward the regions of bump states ($\sigma^{\mathrm{eff}}  > 0$).

\subsection{Adaptive transitions from  negative to positive effective coupling values}
\label{sec:negative-positive}

\begin{figure}[h]
\begin{center}
\includegraphics[height=0.84\textwidth]{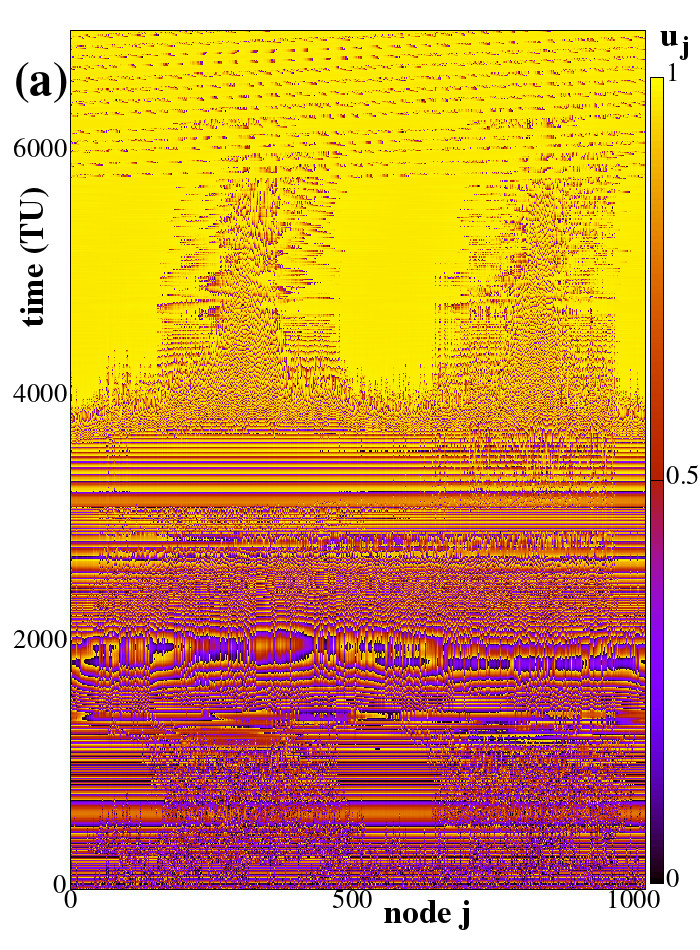}\hspace{3.1mm}
\includegraphics[height=0.89\textwidth]{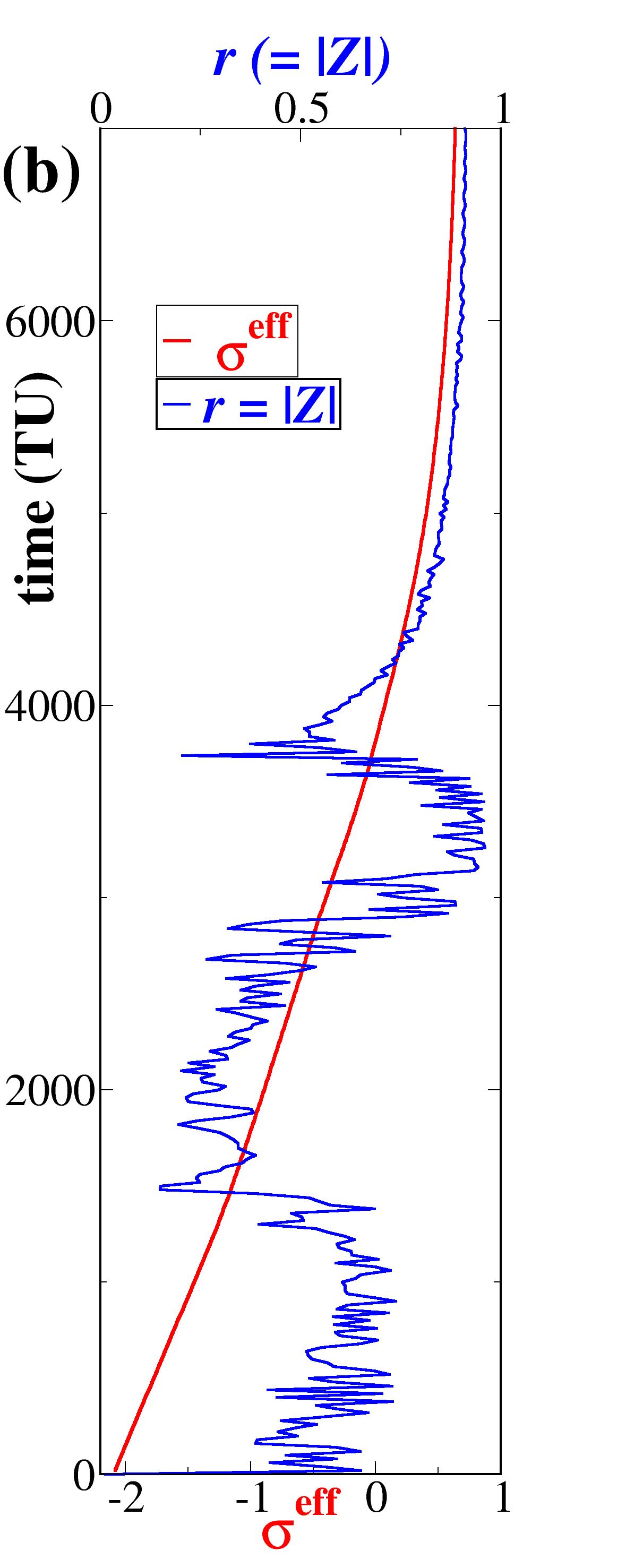}
\end{center}
\caption{\label{fig02}
(Color online) (a) Spacetime plot of the LIF network under adaptive dynamics and (b) the average effective 
coupling strength, $\sigma^{\mathrm{eff}}$, (in red color)
 and the Kuramoto order parameter, $r$,
(in blue color) as a function of time. Note that $\sigma^{\mathrm{eff}}$
 increases, starting from -2 at $t=0$ and approaches asymptotically
the $c_u (=+0.7)$ value. Other parameters are as in Fig.~\ref{fig01}.
All simulations start from random, uniformly distributed initial potentials.
}
\end{figure}

For studying the inverse scenario, we need to start from negative effective couplings 
where chimera states prevail and, through the
coupling evolution, to end in a positive coupling regime where the bump states dominate.
To this end, we start from
an initial state with equal coupling strength, $ \sigma_{jk}(t=0)=-3.0$,
and using as a control parameter $c_u=+0.7$, the effective initial weights take the value
 $ \sigma^{\mathrm{eff}}_{jk}(t=0)=-2.1$. This negative, effective couplings place the system in the 
regime of chimeras at $t=0$. Setting as before the Oja parameter  to the value $\alpha =1$,
 the weights will finally converge asymptotically 
to $\sigma_{jk}(t\to\infty)=1/{\alpha}=1$ and, therefore, the asymptotic effective couplings
will converge  
to $\sigma^{\mathrm{eff}}_{jk}(t\to\infty)=c_u/{\alpha}=+0.7$, which is located in the parameter range
where bumps are found. This scenario is exactly the opposite to the one in Sec.~\ref{sec:positive-negative}
and is demonstrated in Fig.~\ref{fig02}.

\par In Fig.~\ref{fig02}a, the spacetime plot of the system is presented. At $t=0$, the system
starts with randomly and uniformly distributed potentials, as testified by the low values of
the Kuramoto order parameter in Fig.~\ref{fig02}b (blue line). Soon after the dynamics has been switched
on, the system develops a two-headed chimera state which dominates up to about 1300 TUs. 
Besides the spacetime plot, the presence of the chimera state is 
also testified by the increase in the value of the Kuramoto
order parameter taking finite values above 0 and below 1. Driven by Eq.~\eqref{eq03c},
the average effective coupling strength, which has started in negative values,
 increases continuously, Fig.~\ref{fig02}b (red line). At around $t\approx 1300$ TU, a relatively 
abrupt transition takes place and the two-headed chimera reduces to an all-incoherent system,
sustained between times $\approx 1300 - 2100$ TU. The Kuramoto order parameter in this time interval
keeps low but finite values with a tendency to increase in time, indicating that 
there can be small, hidden synchronized domains in the system. 

\par After $t\approx 2100$ TU,
the transition to a one-headed chimera becomes apparent, since synchronized domains appear
 on the left and the right of the large incoherent domain of the chimera, while the values
of $r$ increase accordingly.  As the size of the synchronous domain increases, the asynchronous
region shrinks and this tendency is captured by the Kuramoto order parameter which
obtains values very close to 1 between time $\approx 3200 - 3600$ TU. Around $t\approx 3800$ TU the
effective coupling strength vanishes to 0 and the nodes become decoupled. The Kuramoto order
parameter decreases without ever reaching 0, because many of the oscillators have been
previously synchronized. As the average coupling strength increases further due to the dynamics (Eq.~\eqref{eq03c})
and acquires positive values, 
the systems enters the bumps regime and forms two active bumps mediated by two subthreshold (yellow)
domains. 

\par Having entered the bumps regime, the size of the incoherent domains decrease as the 
effective coupling grows and so does the Kuramoto order parameter. In these parameter ranges
the potential of the subthreshold elements take all similar values, slightly below $u_{\rm th}$
and the Kuramoto order parameter counts them as synchronous. Finally, as the  effective coupling
increases towards its limiting value, $\sigma^{\mathrm{eff}}_{jk}(t\to\infty)=c_u/{\alpha}=0.7$,
the active regions break into a number of small active traveling packets/bumps. The Kuramoto order parameter
tends to increase with time, since the size of the active (incoherent) packets shrinks as the 
coupling range increases
toward its asymptotic value and the subthreshold regions dominate in the system.

\par Overall, in an adaptive network following the Oja rule for the evolution of link weights,
when the effective weights evolve from negative to positive values the system transits from chimera 
to bump states, but the route is not in a one-to-one correspondence with the opposite 
case, where the links evolve from positive to negative values. Similar results can be
shown for different parameter values, see e.g., in the Supplementary Material (Sec. I),
for coupling ranges $R=200$.
In the next section, we will
question the role of the constant $\tau_\sigma$ which controls the relative speed of the 
coupling evolution process with respect to the potential evolution process.

\section{The role of the time scale $\tau_{\sigma}$}
\label{sec:time-scales}

The role of the time scales can be crucial in the evolution of dynamical systems, because
slow time scales give the system the opportunity to spend time near certain states and
develop their characteristic features, whereas fast time scales drive the system directly
to the asymptotic steady state, without allowing it to develop the characteristics
of intermediate scales.

\par In this section, we demonstrate 
 the evolution of the system dynamics when $\tau_\sigma =2$,
i.e., two and a half orders of magnitude lower than the ones in Sec.~\ref{sec:effects-adaptive}.
All other parameter stay the same. Following the analysis in
Sec.~\ref{sec:effects-adaptive}, we study the cases: a) the system starts with 
randomly and uniformly distributed initial potentials and with  $\sigma^{\mathrm{eff}}_{jk}(t=0)= -2.1$
tending asymptotically to $\sigma^{\mathrm{eff}}_{jk}(t\to \infty)= +0.7$ and b) the system starts 
 with  $\sigma^{\mathrm{eff}}_{jk}(t=0)= +2.1$
and ends asymptotically to $\sigma^{\mathrm{eff}}_{jk}(t\to \infty)= -0.7$. The results are presented
in the following two figures.

\begin{figure}[h]
\includegraphics[width=0.49\textwidth]{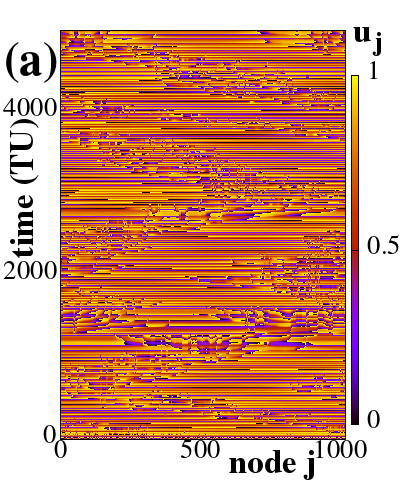}\hspace{2mm}
\includegraphics[width=0.49\textwidth]{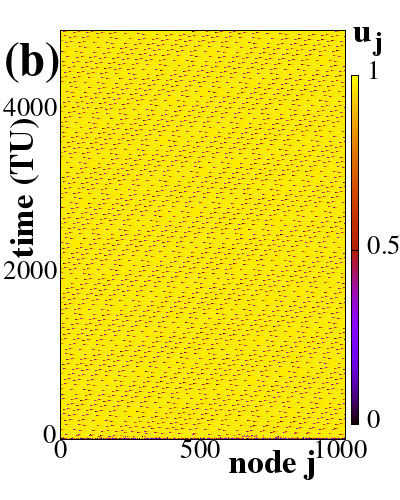}\hspace{2mm}
\caption{\label{fig03}
(Color online)  Spacetime plots of the LIF network under adaptive dynamics.
Here the coupling strengths develop in time scales similar to the potentials,
 $\tau_\sigma=2$. (a) $\sigma^{\mathrm{eff}}$
 decreases, starting from +2.1 at $t=0$ and approaches asymptotically
the $c_u (=-0.7)$ value and (b) $\sigma^{\mathrm{eff}}$
 increases, starting from -2.1 at $t=0$ and approaches asymptotically
the $c_u (=+0.7)$ value.  All other parameters are as in Fig.~\ref{fig01}.
All simulations start from random, uniformly distributed initial potentials.
}
\end{figure}
\par
\begin{figure}[h]
\begin{center}
\includegraphics[width=0.8\textwidth]{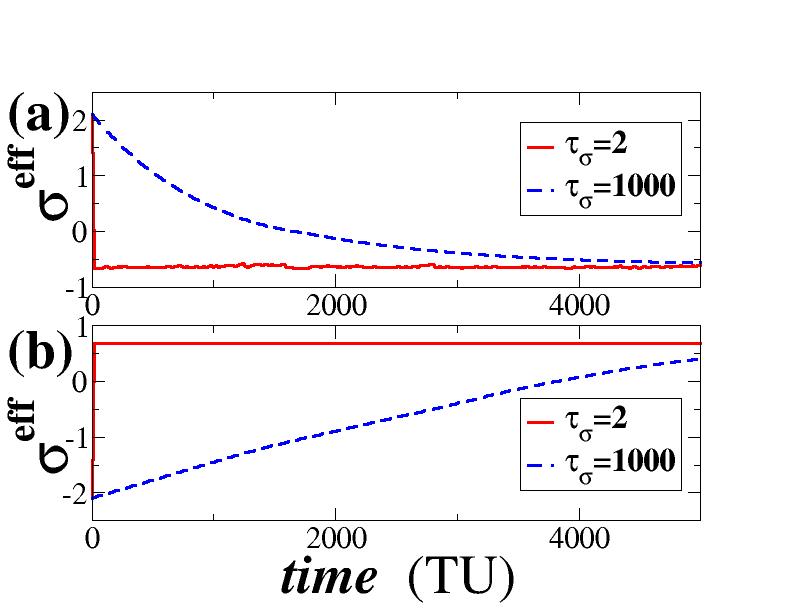}
\end{center}
\caption{\label{fig04}
(Color online)
The average effective 
coupling strength $\sigma^{\mathrm{eff}}$  of the LIF network under adaptive dynamics.
Here the time scale governing the coupling dynamics is $\tau_\sigma=2$,
similar to the ones in the equations for the potentials. 
(a) $\sigma^{\mathrm{eff}}$
 decreases, starting from +2.1 at $t=0$ and approaches asymptotically
the $c_u (=-0.7)$ value and (b) $\sigma^{\mathrm{eff}}$
 increases, starting from -2.1 at $t=0$ and approaches asymptotically
the $c_u (=+0.7)$ value.  Other parameters are as in Fig.~\ref{fig01}.
}
\end{figure}

 In Fig.~\ref{fig03}a, the network that has started from $\sigma^{\mathrm{eff}}_{jk}=-2.1$
due to the fast time scale of Eq.~\eqref{eq03c}, ($\tau_\sigma=2$),
almost immediately reaches its asymptotic state, without demonstrating
intermediate behavior and transitions. In fact, the state reached in Fig.~\ref{fig03}a
is very similar to the one reached in Fig.~\ref{fig01}a after about 6500 TU.
Similarly, in Fig.~\ref{fig03}b, the network has started from $\sigma^{\mathrm{eff}}_{jk}=+2.1$ and
almost immediately reaches its asymptotic traveling bumps state. Here also,
the behavior of the system shows similar characteristics as in Fig.~\ref{fig02}a,
after approximately 6500 TU.
These immediate collapses to the asymptotic states, without passing from
intermediate partially organized states, are attributed to the fast 
time scaling governing the adaptive coupling dynamics.
The evolution of the average effective coupling ranges
confirms these observations, see Fig.~\ref{fig04}.  

\par In Fig.~\ref{fig04}a, we plot using red solid line the average effective coupling
strength for fast evolving couplings, using $\tau_\sigma=2$. We note that the
system almost immediately reaches its asymptotic state, $\sigma^{\mathrm{eff}}=c_u/\alpha =-0.7$.
For comparison, we also plot using blue dotted line the average effective coupling
strength for the slowly evolving couplings shown in the previous section, where $\tau_\sigma=1000$.
In the latter case, the system reaches its asymptotic state only after 4000-5000 TU.
Following a similar and even faster scenario, is the case where the system starts
in the negative coupling ranges, $\sigma_{jk}^{\mathrm{eff}} (t=0)=-2.1$, and evolves toward positive ones,
see Fig.~\ref{fig04}b. Comparison of the evolution under low time scales,  
$\tau_\sigma=2$ (red solid line), with high ones,  $\tau_\sigma=1000$ (blue dotted line), 
also reveals the fast, almost immediate approach to the asymptotic bump state for $\tau_\sigma=2$.

\par
\begin{figure}[h]
\begin{center}
\includegraphics[width=0.7\textwidth]{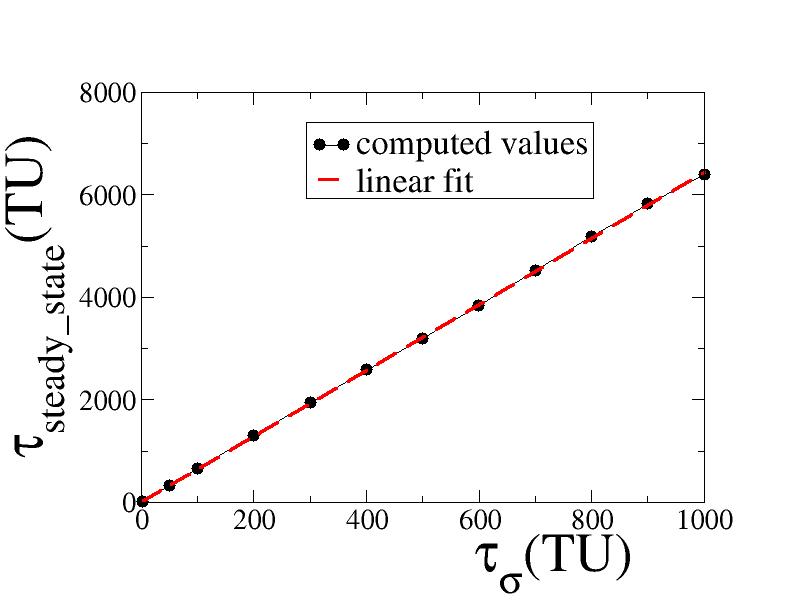}
\end{center}
\caption{\label{fig05}
(Color online) 
Plot of the time $\tau_{\rm steady\_state}$ that the network needs to reach its asymptotic
steady state as a function of the timescale $\tau_{\sigma}$ governing the evolution of
the coupling weights. The actual $\tau_{\rm steady\_state}$ is calculated using a tolerance distance
$\epsilon=0.1$ away from the steady state. Other parameters are $\alpha=1$ and $c_u=0.7$.
Initial coupling values are $\sigma^{\mathrm{eff}}_{jk}(t=0)=c_u\> \sigma_{jk}(t=0)=-2.1$
and final $\sigma^{\mathrm{eff}}_{jk}(t\to \infty)=c_u/\alpha =0.7$. Therefore, the actual $\tau_{\rm steady\_state}$
values are here obtained when $\sigma^{\mathrm{eff}}_{jk}(t)=c_u/\alpha -\epsilon =0.6$.
Other parameters are as in Fig.~\ref{fig01}.
}
\end{figure}

 To have an idea of the time that the system needs to reach its final state, which
in the case of the Hebb-Oja rule is achieved when $\left< \sigma_{jk}\right> =1/\alpha =1$
(or $\sigma^{\mathrm{eff}}(t \to \infty) =c_u/\alpha =0.7$), we performed
numerical simulations using different timescale factors, $1 \le \tau_\sigma \le 1000$,
and calculated the time $\tau_{\rm steady\_state}$ that the system needs until it reaches
its final coupling configuration. In fact, because the system may reach the final
coupling configuration at asymptotically large times, we allow a deviation of 
order $\epsilon$ from the asymptotic state and calculate the time needed to reach an average effective
coupling weight equal to $c_u/\alpha -\epsilon$. In Fig.~\ref{fig05}, we plot $\tau_{\rm steady\_state}$
as a function of $\tau_{\sigma}$, for  $1 \le \tau_\sigma \le 1000$. For this example we
use as parameters $c_u=0.7$ and $\alpha=1$. For all simulations the initial coupling strengths
are $\sigma^{\mathrm{eff}}_{jk}(t=0)= c_u\> \sigma_{jk}(t=0)=-2.1$.
 To avoid waiting infinitely long times
to reach the asymptotics, we estimate the time $\tau_{\rm steady\_state}$ using a tolerance
$\epsilon =0.1$, or  the actual $\tau_{\rm steady\_state}$
values are recorded when $\sigma^{\mathrm{eff}}_{jk}(t)=c_u/\alpha -\epsilon =0.6$. The $\tau_{\rm steady\_state}$ values
are computed with an error of the order of $\pm 10$ because the data are recorded every 10 TU.

Figure~\ref{fig05} shows that the time to reach the asymptotic coupling state increases
linearly with the timescale governing the coupling evolution, namely: 
$\tau_{\rm steady\_state}=
A\> \tau_{\sigma} +B $. Note that the values of the fitted parameters $A$ and $B$
depend on the tolerance $\epsilon$ and on other system parameters, such as $c_u$, $\alpha$,
and others. For the given parameter values,
linear fitting to the data allows to estimate $A(\epsilon)$ and 
$B(\epsilon)$. For $\epsilon=0.1$ the fitting gives $A=6.4$ and $B=5.3$. 
The following argument allows to estimate the value of $B$: From Eq.~\eqref{eq03c}
it is expected that if $\tau_{\sigma}=0$ this has an equivalent effect as the time
derivative $d\sigma_{jk}/dt $
 being equal to zero. In both cases the system is found at the steady state and therefore
 $\tau_{\sigma}=0$ entails that $\tau_{\rm steady\_state}=0$. Inserting this point,
$(\tau_{\sigma},\tau_{\rm steady\_state})=(0,0)$, to the linear relation we conclude that
the fitting parameter $B=0$. This estimated value lines up with the computed value, $B=5.3$,
because all data points are recorded with an error of $\pm 10$, as explained in the
previous paragraph. Similar behaviour is obtained when we simulate the opposite process,
namely, starting from initial coupling strengths
 $\sigma^{\mathrm{eff}}_{jk}(t=0)= c_u\> \sigma_{jk}(t=0)=2.1$ (in the bump state regime)
we tend asymptotically toward  the final chimera state regime with
 $\sigma^{\mathrm{eff}}_{jk} \to c_u =-0.7$ (see results and discussion in the Supplementary
Material, Sec. II).

\par The overall conclusions on the effects of the parameter $\tau_\sigma$ is that when the
time scales of the coupling evolution are similar to the ones of the potential evolution,
the network crosses very fast its intermediate states and reaches almost immediately
its asymptotic state. In neural systems this similarity of the time scales is achieved under 
short-term plasticity conditions.  To the contrary, when the coupling evolution takes place
 much slower than the potential, the system has the time to visit states with different
synchronization properties and demonstrates interesting transitions between these states.
In neural systems these last transitions are achieved under long-term potentiation conditions
and could be associated with learning processes.

\section{Discussion on the form of the coupling matrix}
\label{sec:discussion}

In this section, we discuss the form of the coupling matrix, $\sigma_{jk}$,
and the distribution of the coupling weights as the adaptive LIF network crosses the 
various synchronization regimes. First, in Sec.~\ref{sec:bumps-chimeras},
 we present the case when the system crosses from
bump to chimera states and then, in Sec.~\ref{sec:chimeras-bumps}, the opposite case.
In the entire section \ref{sec:discussion}, the effective form of the coupling
matrix, $\sigma^{\mathrm{eff}}_{jk}$, will be used, because it accounts for the overall
$(\, c_u\> \sigma_{jk}\, )$ coupling weights in Eq.~\eqref{eq03a}.

\subsection{Coupling distributions when crossing from bumps to chimera states}
\label{sec:bumps-chimeras}

\par
\begin{figure}[h]
\begin{center}
\includegraphics[height=0.41\textwidth]{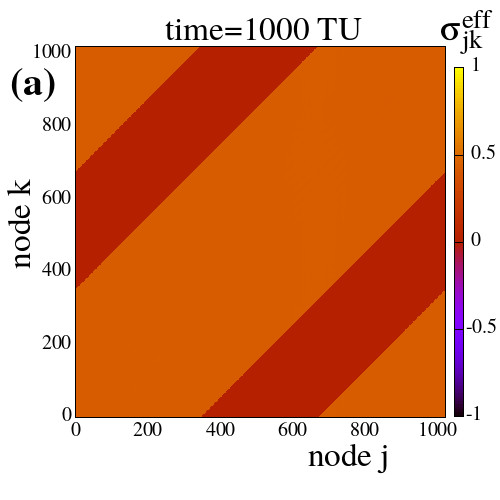}\hspace{2mm}
\includegraphics[height=0.41\textwidth]{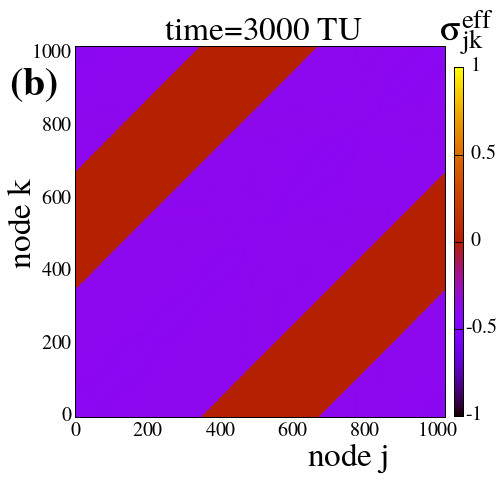}\\[2mm]
\includegraphics[height=0.41\textwidth]{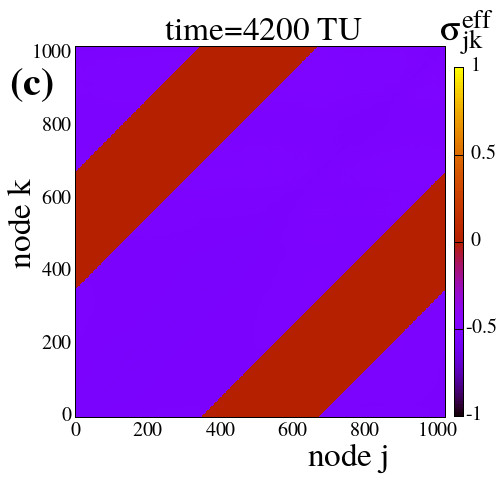}\hspace{2mm}
\includegraphics[height=0.41\textwidth]{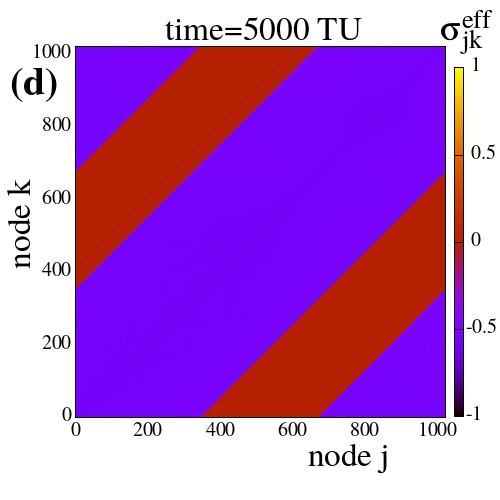}
\end{center}
\caption{\label{fig06}
(Color online) 
Color representation of the effective coupling matrix at four different time instances.
The coupling weights, $\sigma^{\mathrm{eff}}_{jk}$,
 decrease with time. At $t=0$, they have started from the value 2.1 and approach asymptotically
the $\sigma^{\mathrm{eff}}_{jk}=c_u/\alpha =-0.7$ value. (Respectively, the coupling weights, $\sigma_{jk}$,
start from -3.0 at $t=0$ and approach asymptotically
the $1/\alpha =1$ value.).   The representative time instances are:
(a) $t=1000$ TU, in the bump state regime, (b) $t= 3000$ TU, in the two-headed chimera regime,
(c)  $t= 4200$ TU, in the single-headed chimera regime, and 
(d)  $t= 5000$ TU, in the chaotic synchronization regime.
All parameters are as in Fig.~\ref{fig01}.
}
\end{figure}

 In Fig.~\ref{fig06}, we plot the profile of the coupling matrix at
four different instances: panel (a) at $t=1000$ TU, when the network is in the bump state, panel (b) at $t= 3000$ TU,
when the network has crossed into the two-headed chimera regime, panel 
(c) at $t= 4200$ TU, in the single-headed chimera regime, and panel
(d) at $t= 5000$, when the system is in the regime where the synchronous and asynchronous regions move
 chaotically around the network.

\begin{figure}[h]
\includegraphics[clip,width=0.8\linewidth, angle=0]{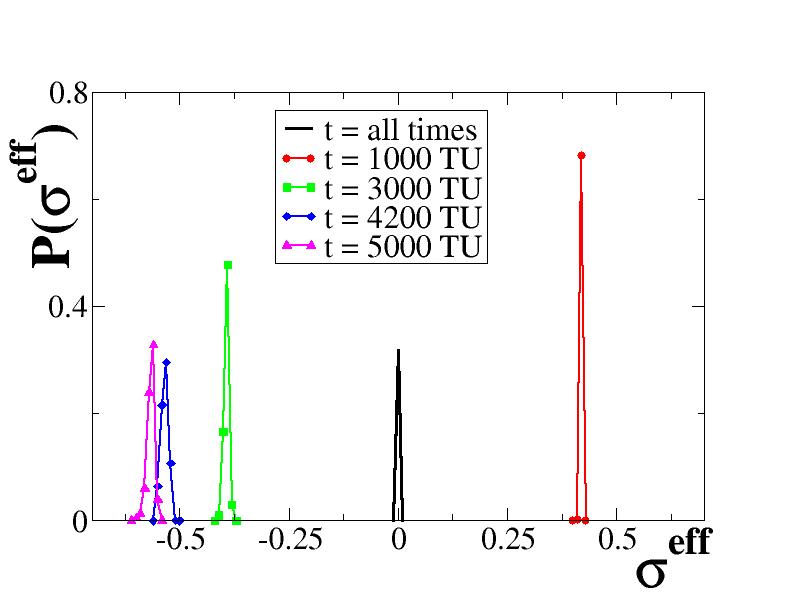}
\caption{\label{fig07} (Color online) 
The distribution functions of the effective coupling strengths, $\sigma^{\mathrm{eff}}_{jk}$, for the four different
instances indicated in Fig.~\ref{fig06}. The red line with solid circles corresponds
to $t=1000$ TU in the bump state regime. The green line with solid squares
corresponds to $t= 3000$ TU, in the two-headed chimera regime.
The blue line with solid diamonds corresponds to  $t= 4200$ TU, in the single-headed chimera regime.
The magenta line with triangles corresponds to  $t= 5000$ TU, in the chaotic synchronization regime.
The black line is common to all time steps and corresponds to the connections with 0 weights.
 All parameters are as in Fig.~\ref{fig06}. 
}
\end{figure}

In all four panels the color variable takes values between [-1, 1] to chromatically compare the changes
in values of the coupling as the system crosses through the various synchronization states. First, the uncoupled
connections are all colored brown-red, which correspond to values $\sigma^{\mathrm{eff}}_{jk}=0$ and is conserved
in all times (and consequently in all four panels). 
In panel (a) (t=1000 TU), the effective weights of the connected nodes take very similar positive
values. This is because the system has started from $\sigma_{jk}=-3.0$, which corresponds
to positive effective values $\sigma^{\mathrm{eff}}_{jk}=c_u\> \sigma_{jk}\approx 2.1$ (given that $c_u=-0.7$) and
during the first 1000 TU did not have enough time to cross into negative effective weights. The almost uniform
orange color in Fig.~\ref{fig06}a shows that all $\sigma^{\mathrm{eff}}_{jk}$(t=1000 TU) values are very close.
This does not contradict the fact that bump states are possible under conditions of identical
coupling strengths as well as identical elements \cite{tsigkri:2018}. The same conclusion is also
obtained if we calculate the distribution of coupling strengths, $P(\sigma^{\mathrm{eff}})$. 
The distribution $P(\sigma^{\mathrm{eff}})$ for panel Fig.~\ref{fig06}(a) is plotted in Fig.~\ref{fig07} with
the red-curve combined with the black-curve. The red-curve accounts for the nonzero elements,
while the black-curve accounts for the null weights. The red-curve designates a very narrow, sharp
distribution with almost identical values as  visually seen in Fig.~\ref{fig06}(a).
The black-curve represents a $\delta$-like function at $\sigma^{\mathrm{eff}} =\sigma=0$ and is common in all graphs since
the position and strength of the null connections do not change in time.

\par For the same system we also plot in Fig.~\ref{fig06}(b) the color-coded profile of the
effective coupling strengths at time t=3000 TU, when the system has entered the two-headed chimera regime. 
The resulting plot shows that the non-zero coupling values have moved to lower values (purple colors),
close to $\sigma^{\mathrm{eff}}_{jk}=-0.4$ (or $\sigma_{jk}=0.5$). The same conclusion is drawn from the corresponding
green distribution (always combined with the black one), in Fig.~\ref{fig07}. 
The distribution of weights spreads slightly but remains basically narrow 
  as it travels to the left. It is interesting to note
that although the network has crossed from positive to negative $\sigma^{\mathrm{eff}}$ values and
from bump states to chimera states, the form of $P(\sigma)$ does
not undergo important modifications except for a slight increase in the width of the
distribution (with corresponding decrease in its height). 

\par The last two panels in Fig.~\ref{fig06} correspond to single-headed chimera states in
panel (c) when the incoherent region moves toward the right, while in panel (d) the incoherent region
wanders erratically, breaks and remerges in time. In both cases the non-zero elements
of the coupling matrix do not show broad distributions, as also confirmed from the
corresponding curves in Fig.~\ref{fig07}. Indeed, the blue curve (together with the
black curve) which corresponds to Fig.~\ref{fig06}(c) has moved toward lower negative
values and shows an increased width, relatively to the red and green ones and so does
the magenta curve which corresponds to Fig.~\ref{fig06}(d), for time $t=5000$ TU.
This last curve approaches closer the asymptotic value of $\sigma^{\mathrm{eff}}_{jk}$, expected to be
equal to $c_u/\alpha =-0.7$ (or $\sigma_{jk}\to 1/\alpha =1$) in our simulations.

\par Overall, the distributions of $\sigma$-values keep a narrow profile
 as the system crosses from bumps to chimera states which slightly widens as the
system enters in the chimera regions. The distributions
show a tendency to shrink again as the coupling strengths slowly tend to their asymptotic values.

\subsection{Coupling distributions when crossing from chimera to bump state regimes}
\label{sec:chimeras-bumps}

 As discussed earlier, to obtain the evolution from chimeras to bump states we just need to change the
control parameter $c_u$ into positive values. Here the value $c_u=+0.7$ is used, while all other
parameters remain the same. With this modifications, $\sigma_{jk}$ start at -3.0 and tend asymptotically
to the value $1/\alpha =1$. Respectively, $\sigma^{\mathrm{eff}}_{jk}$ start at -2.1 (setting the system in the chimera regime)
 and tend asymptotically to $c_u/\alpha=+0.7$ (setting the system in the bumps regime).
\par
\begin{figure}[h]
\begin{center}
\includegraphics[height=0.41\textwidth]{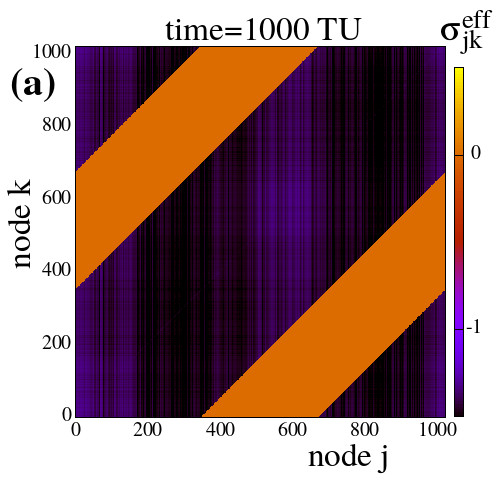}\hspace{2mm}
\includegraphics[height=0.41\textwidth]{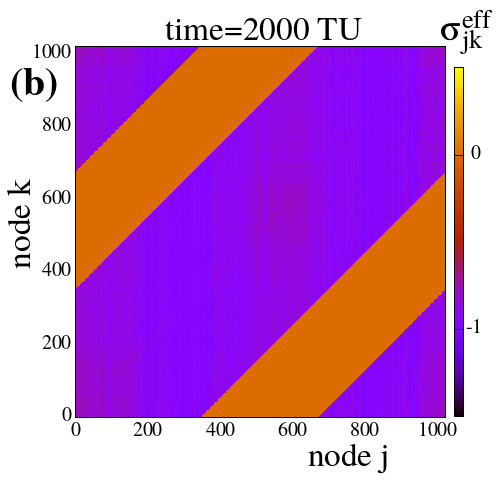}\\[2mm]
\includegraphics[height=0.41\textwidth]{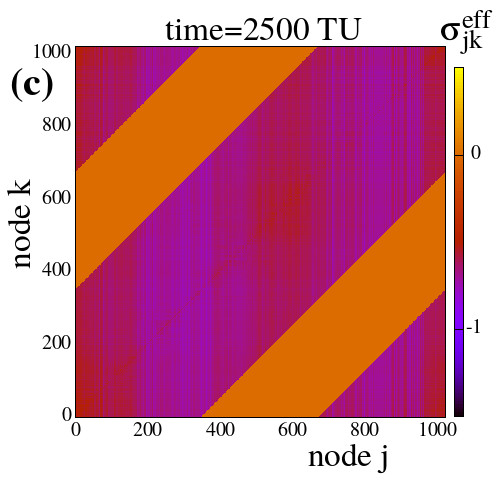}\hspace{2mm}
\includegraphics[height=0.41\textwidth]{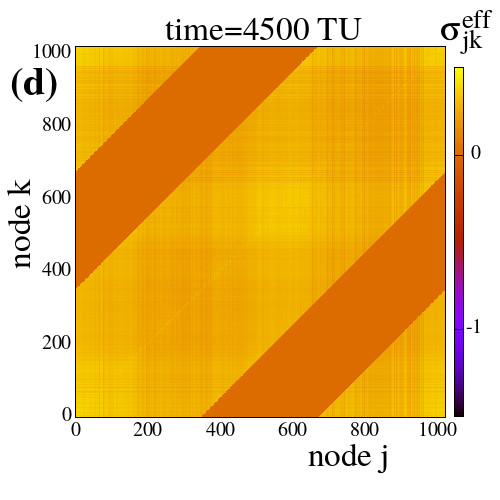}
\end{center}
\caption{\label{fig08}
(Color online) 
Color representation of the effective coupling matrix, $\sigma^{\mathrm{eff}}_{jk}$, at four representative time instances.
The effective coupling weights
start from -2.1 at $t=0$ and approach asymptotically
the $c_u (=+0.7)$ value. The representative time instances are:
(a) $t=1000$ in the two-headed chimera state, (b) $t= 2000$, in the
transition region between  two-headed and single-headed chimeras,
(c)  $t= 2500$, in the single-headed chimera regime, and 
(d)  $t= 4500$, in the two-bumps state.
All parameters are as in Fig.~\ref{fig02}.
}
\end{figure}

 To explore this transition,
we plot in Fig.~\ref{fig08} four representative color-coded effective coupling profiles
for times: $t=1000$ TU in panel (a)  corresponding to the two-headed chimera state, 
 $t= 2000$ TU in panel (b) corresponding to the
transition region between two-headed and single-headed  chimeras,
$t= 2500$ TU in panel (c)  corresponding to the single-headed chimera regime, and 
$t= 4500$ TU  in panel (d) with the system in the two-bumps state.

\par
\begin{figure}[h]
\includegraphics[clip,width=0.8\linewidth, angle=0]{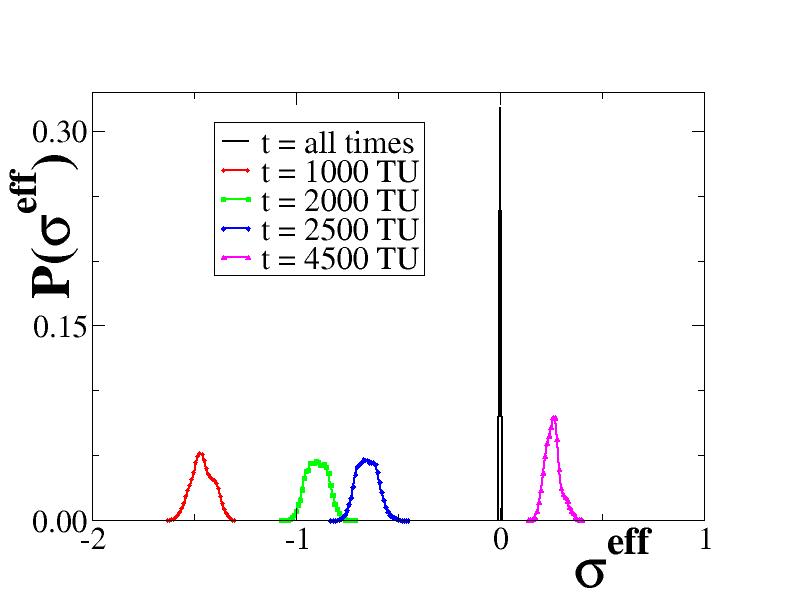}
\caption{\label{fig09} (Color online)
The distribution functions of the  effective coupling strengths, $\sigma^{\mathrm{eff}}_{jk}$, for the four different
instances indicated in Fig.~\ref{fig08}. 
The red line (solid circles) corresponds
to $t=1000$ TU in the two-headed chimera state. The green line (solid squares)
corresponds to $t= 2000$ TU, in the transition region between two-headed and one-headed chimera states.
The blue line (solid diamonds) corresponds to  $t= 2500$ TU, in the one-headed chimera regime.
The magenta line (solid triangles) corresponds to  $t= 4500$, in the two-bumps state.
The black line is common to all time steps and corresponds to the connections with 0 weights.
All parameters are as in Fig.~\ref{fig08}. 
}
\end{figure}

 For the same system we also plot the distribution $P(\sigma^{\mathrm{eff}})$ of coupling strengths in the four 
instances presented in Fig.~\ref{fig08}. The results are shown in Fig.~\ref{fig09}. Visual inspection
of the four panels in Fig.~\ref{fig08} already demonstrates that the coupling strengths are not 
constant or uniform but they present considerable spreading of values. This is further illustrated in
Fig.~\ref{fig09}, where the distributions are not narrow any more, but acquire a considerable variance
around the mean.

\par More specifically, in Fig.~\ref{fig08}(a), the non-zero effective coupling strengths are colored
black or dark-purple, around values $\sigma^{\mathrm{eff}}_{jk}\approx -1.5$. As noted before, the coloring
is not uniform demonstrating the presence of different coupling strengths in the system. The linear (vertical in this
case) formations indicate that certain neurons provide uniform coupling strengths to all other coupled neurons
in the system. The different coupling values are evidenced in the red colored curve in Fig.~\ref{fig09}
which together with the black curve account for the full distribution of weights in  
Fig.~\ref{fig08}(a). The situation has not significantly changed as we enter from the two-headed
chimera (panel (a)) to the one-headed chimera (panel (b)). In fact, the color coded $\sigma^{\mathrm{eff}}$ profile in
Fig.~\ref{fig08}b demonstrates the overall increase of the coupling strengths (lighter purple colors), while the linear
formations and the overall
structure do not significantly change. Respectively, the green curve in Fig.~\ref{fig09} 
demonstrates the increase of the effective couplings without important variations in the coupling strength
distributions. 

\par As time increases, at $t=2500$ TU, the system has entered the one-chimera regime, while there
are some non-significant changes in the formation in Fig.~\ref{fig08}c. The corresponding blue
curve in Fig.~\ref{fig09} demonstrates mostly the overall increase of the coupling strengths 
toward values $\sigma^{\mathrm{eff}}_{jk}\approx -0.7$. Finally, in times of the order
of $t=4500$ TU (Fig.~\ref{fig08}d) the coupling values have crossed over into positive values, 
entering the regime of bump states. The formations in panel (d) still
persist, however, the corresponding magenta curve in Fig.~\ref{fig09}, which is now beyond
the $x=0$ axis, shows a tendency to shrink, following the Oja dynamics where, asymptotically, all coupling
weights
$\sigma_{jk} \to 1/\alpha =1.0$, or equivalently $\sigma^{\mathrm{eff}}_{jk} \to c_u/\alpha =0.7$.

\par Overall, for adaptive dynamics leading from the chimera to the bump states regimes,
the coupling values show a considerable spreading (compared to the opposite case),
while the distributions shrink as the coupling weights approach the asymptotic value $\sigma_{jk}\to 1/\alpha$.

\section{Conclusions and Open Problems }
\label{sec:conclusions}
\par In this study we have considered the adaptive co-evolution of the potentials and the
coupling strengths in a network of coupled LIF units following the Hebb-Oja rule for the updating
of the coupling weights. We showed that slow evolution of the coupling weights allows the system to cross
from negative to positive couplings (and the opposite) and to go through various
partial synchronization regimes. We also showed evidence that fast evolution of the coupling
strengths leads directly to the asymptotic states, without allowing the system to develop
intermediate synchronization states. This study demonstrates the importance of the time scales
of the various subprocesses in the overall evolution of a dynamical system. Regarding
the use of quantitative measures for monitoring the network, we note
that the Kuramoto order parameter is more sensitive and demonstrates abrupt transitions as the system crosses
the different synchronization regimes, while in the average coupling strength the transitions
are smooth. 

\par In Figs.~\ref{fig01}a and ~\ref{fig02}a we have noted that
 double chimera states emerge very fast before or after the transition near $\sigma^{\mathrm{eff}}=0$. It would be interesting
to investigate the possibility of the presence of chimera states of higher multiplicity if we consider
even lower time scales for the evolution of the couplings in Eq.~\eqref{eq03c}, e.g. $\tau_{\sigma}=10^4-10^6$.
Similarly, for the case of bump states with many (more than two) active/inactive domains.
Providing the system with adequate time to traverse and explore its range of synchronization states 
may facilitate the detection of finer and finer transitions between distinct synchronization regimes.

\par The gradual change of the system's features as the coupling parameters ``slide'', reminds us of the
continuation approaches used in dynamical systems. Namely, when calculating the system features
for a certain parameter value $p$, we initialize the system using the final steady state values of the system
at a nearby parameter value $p-dp$. This special initialization is called ``continuation'' and
is used to save computational time because the system is initialized near its steady state
(under the assumption that nearby parameter values lead to nearby steady states). The present
co-evolution of the coupling weights operate in 
the a form of a continuation process: as the couplings
evolve slowly, the steady states are drawn by them to cross the various synchronization regimes
before reaching the final asymptotic state.

\par Besides the LIF model, it would be interesting to investigate the effects of (slow and fast)
weight adaptation in other dynamical systems, such as the FitzHugh Nagumo, the Hindmarsh-Rose,
the Van der Pol and the Kuramoto networks of coupled nonlinear oscillators.

\section*{Acknowledgments}
The authors would like to thank Prof. W. Li for helpful discussions.
This work was 
supported by computational time granted from the Greek Research \& Technology Network (GRNET)
in the National High Performance Computing HPC facility - ARIS - under project ID PR014004.

\section*{Data Availability Statement}
Data is available upon request from the authors.
%
%
\bibliography{./provata2025-adaptive.bib}

\end{document}